\documentclass[modern,12pt]{aastex63}
\makeatletter
\let\frontmatter@title@above=\relax
\makeatother
\usepackage{times}
\usepackage{geometry}
\usepackage{titlesec}
\titleformat{\section}
  {\normalfont\fontsize{14}{14}\bfseries}{\thesection}{1em}{}
\usepackage[utf8]{inputenc}
\usepackage{enumitem,amssymb}
\usepackage{ragged2e}
\usepackage{graphicx}
\usepackage{floatrow}
\usepackage{color}
\usepackage{verbatim}
\newlist{thematic}{itemize}{8}
\setlist[thematic]{label=$\square$}
\usepackage{pifont}
\usepackage{orcidlink}
\begin{document}
\pagenumbering{gobble}
\RaggedRight
\noindent {\fontsize{16}{20} \selectfont White Paper for the 2024 Solar \& Space Physics Decadal Survey}
% \vspace{-0.4cm}
% \begin{center}

\centering
\vspace{0.5cm}
{\fontsize{20}{32}\selectfont Advancing Theory and Modeling Efforts in Heliophysics}
\vspace{0.5cm}

%\textit{\fontsize{16}{20}\selectfont New Frontiers with a Next-Generation Solar Radio Facility}
% \end{center}

%\vspace{0.3cm}

\normalsize

\justifying

%\begin{figure}[!ht]
%\floatbox[{\capbeside\thisfloatsetup{capbesideposition={right,top},capbesidewidth=4cm}}]{figure}[\FBwidth]
%{\includegraphics[width=1.0\textwidth]{title_image.pdf}}
%\end{figure}

\noindent \textbf{Main Category:} Basic research \\ \textbf{Subcategories 1:} Fundamental space plasma theory, \\ \textbf{Subcategories 2:} Basic research that is driven by space weather user needs \\

\noindent \textbf{Principal Author:} \\
Fan Guo\orcidlink{0000-0003-4315-3755}, Los Alamos National Laboratory \\
Email: \href{mailto:guofan@lanl.gov}{guofan@lanl.gov} \\
\textbf{Co-authors}\\
Spiro Antiochos\orcidlink{0000-0003-0176-4312}(U. of Michigan), Paul Cassak\orcidlink{0000-0002-5938-1050}(West Virginia University),
Bin Chen\orcidlink{0000-0002-0660-3350}(NJIT), Xiaohang Chen\orcidlink{0000-0003-2865-1772}(U. of Arizona),
Chuanfei Dong\orcidlink{0000-0002-8990-094X}(PPPL/Princeton), Cooper~Downs\orcidlink{0000-0003-1759-4354}(Predictive Science Inc.), 
Joe Giacalone\orcidlink{0000-0002-0850-4233}(U. of Arizona),
Colby~C.~Haggerty\orcidlink{0000-0002-2160-7288}(U. of Hawaii),
Hantao Ji\orcidlink{0000-0001-9600-9963}(PPPL/Princeton),
Judith Karpen\orcidlink{0000-0002-6975-5642}(NASA/GSFC), James Klimchuk\orcidlink{0000-0003-2255-0305}(NASA/GSFC), Wen Li\orcidlink{0000-0003-3495-4550}(Boston University), Xiaocan Li\orcidlink{0000-0001-5278-8029}(Dartmouth College), Mitsuo Oka\orcidlink{0000-0003-2191-1025}{(U. of California Berkeley)}, Katharine K. Reeves\orcidlink{0000-0002-6903-6832}(Harvard Smithsonian CfA), Marc Swisdak\orcidlink{0000-0002-5435-3544}(U. of Maryland), and
Weichao Tu\orcidlink{0000-0003-4547-3269}(West Virginia University)

\vspace{0.2cm}

\noindent \textbf{Synopsis} \\
Heliophysics theory and modeling build understanding from fundamental principles to motivate, interpret, and predict observations. Together with observational analysis, they constitute a comprehensive scientific program in heliophysics. As observations and data analysis become increasingly detailed, it is critical that theory and modeling develop more quantitative predictions and iterate with observations. Advanced theory and modeling can inspire and greatly improve the design of new instruments and increase their chance of success. In addition, in order to build physics-based space weather forecast models, it is important to keep developing and testing new theories, and maintaining constant communications with theory and modeling. Maintaining a sustainable effort in theory and modeling is critically important to heliophysics. We recommend that all funding agencies join forces and consider expanding current and creating new theory and modeling programs--especially, 1. NASA should restore the HTMS program to its original support level to meet the critical needs of heliophysics science; 2. a Strategic Research Model program needs to be created to support model development for next-generation basic research codes; 3. new programs must be created for addressing mission-critical theory and modeling needs; and 4. enhanced programs are urgently required for training the next generation of theorists and modelers.

%\end{abstract}

\thispagestyle{empty}
%\newpage

\vspace{5.0cm}

\section{Introduction}\label{sec:intro}

\pagenumbering{arabic}
\setcounter{page}{1}

Theory and modeling study heliophysical processes from fundamental principles to understand, interpret, and predict observations. Together with observational analysis, they constitute a comprehensive scientific program in heliophysics.  %Theoretical investigations shed light on unexplained phenomena, build deeper physical insight, and predict new and unnoticed phenomena. 
In response to observational discoveries, new theories are developed and tested. Based on indirect observations at the time, Parker theoretically predicted a supersonic solar wind \citep{Parker1958,Parker1999}, which is directly related to the birth of heliophysics. A complete understanding of heliophysics processes relies on close collaborations between theory/modeling and observation/data analysis. Through such integrated efforts, theory/modeling motivates further observational analysis, data-theory comparison, and new missions/infrastructures. As observations and data analysis become increasingly detailed, it is critical that theory/modeling provides quantitative predictions and iterate with observations for further progress. For example, recent observations by Parker Solar Probe and other spacecraft have shown the prevalence of switchbacks in the inner heliosphere \citep{Bale2019,Kasper2019}, leading to a new set of theories proposed to explain the observations and increasingly stringent tests for those theories \citep[e.g.,][]{Squire2020,Zank2020,Ruffolo2020,Drake2021,Schwadron2021}. In addition, advances in physics-based space weather models depend critically on the knowledge gained from basic theory/modeling, and our ability to transition this knowledge into those models. Therefore, maintaining a sustainable effort in theory and modeling is essential to both basic and applied heliophysics. 

Heliophysical plasma systems host a rich set of physical phenomena and are often the only places that allow detailed remote sensing and in situ observations. Many plasma processes, including magnetic reconnection and collisionless shocks, were discovered or proposed to explain heliophysical phenomena \citep[e.g.,][]{Parker1957,Sweet1958,Blandford1987}. Theoretical understanding of these phenomena has far-reaching impacts on our insight into basic plasma physics, laboratory experiments, and planetary and astrophysical plasmas. An incomplete list of processes (with corresponding white papers) includes dynamos, the acceleration and transport of energetic particles, magnetic reconnection \citep{Ji2022,Chen2022}, collisionless shocks \citep{Goodrich2022}, turbulence and waves in fluid and kinetic scales, wave-particle interaction \citep{Thorne2010}, plasma heating and energy dissipation \citep{Haggerty2022} and effects of partially ionized plasmas. These processes underlie many important heliophysical phenomena, including solar and heliospheric magnetic fields, energetic particles and radiation \citep{Shih2022,Chen2022,Turner2022,Kollmann2022}, coronal and solar wind heating \citep{Klimchuk2022, Mondal2022}, the origin of solar eruptions and magnetospheric storms and substorms \citep{Chen2022,Green2022}, the shape and size of the heliosphere, ionospheric outflow, and the impact of solar activity on planetary magnetospheres and atmospheres in our solar system \citep{Cohen2022,Crary2022} and beyond \citep{Garcia-Sage2022}. %Here we highlight an (incomplete) list of theoretical areas that need significant support and continuous development and note some relevant white papers. These topics are crucial to understanding basic heliophysics and to build sophisticated space weather models.

\begin{comment}
\begin{itemize}
\item  The Acceleration of Energetic Particles \citep{Shih2022,Chen2022}
\item Magnetic Reconnection \citep{Ji2022,Chen2022}
\item Collisionless Shocks 
\item Coronal Heating \citep{Klimchuk2022, Mondal2022}
\item Plasma Heating and Energy Dissipation \citep{Haggerty2022}
\item Energetic Particle Confinement and Transport
\item Partially Ionized Plasmas
\item Plasma Turbulence 
\item Wave-Particle Interaction 
\item Solar Eruptions
\item Global Modeling of the heliosphere
\item Solar Dynamo
\item Interface and Shear Flow Instabilities
\item Earth's Magnetosphere and Comparative Magnetospheres \citep{Garcia-Sage2022}
\item Terrestrial and Planetary Storms and Substorms \citep{Green2022}
\end{itemize}
\end{comment}

\section{The Importance of theory/modeling to observations and instruments}

The exercise of theory/modeling leads to a significantly deeper understanding of observations, identifies gaps in our understanding, and motivates new observations, missions, and infrastructures. We point out that collaborations between theory and observations in recent NASA missions, such as IBEX, MMS, Van Allen Probes, THEMIS, MESSENGER, MAVEN, Juno, IRIS, SDO, Hinode, and Parker Solar Probe, as well as ground-based facilities such as EOVSA and DKIST, have led to major progress in our understanding of key heliophysics phenomena. Current and future space and ground-based missions and infrastructures would benefit greatly from strong theory/modeling participation in augmented research programs and science teams, from the design phase through post-launch data acquisition. In particular, comparing  theory/model predictions with the ground truth of observations can resolve contentious issues and point to new directions for both observations and theory. Figure 1 shows a model-data comparison for the solar flare on 2017 September 10, which was observed by multiple spacecraft/facilities including EOVSA, Hinode, RHESSI, and SDO. The EOVSA observation measures the energetic electron density and magnetic field strength \citep{Chen2020,Fleishman2020}. The MHD model agrees well with the flare dynamics and the distribution of the magnetic field strength, and the EOVSA radio image is consistent with an MHD-particle-radiation model \citep{Chen2020,Li2022}, suggesting that flare current-sheet reconnection and looptop sources are crucial accelerators for energetic electrons. Combining advances in theory/modeling of particle acceleration and flare dynamics and next-generation radio imaging spectropolarimetry, as would be provided by FASR \citep{Chen2022}, will further demystify the role of reconnection and associated particle acceleration processes in flares. 

\begin{figure}[!ht]
%\floatbox[{\capbeside\thisfloatsetup{capbesideposition={right,top},capbesidewidth=4cm}}]{figure}[\FBwidth]
\vspace{-0.3cm}
{\includegraphics[width=0.9\textwidth]{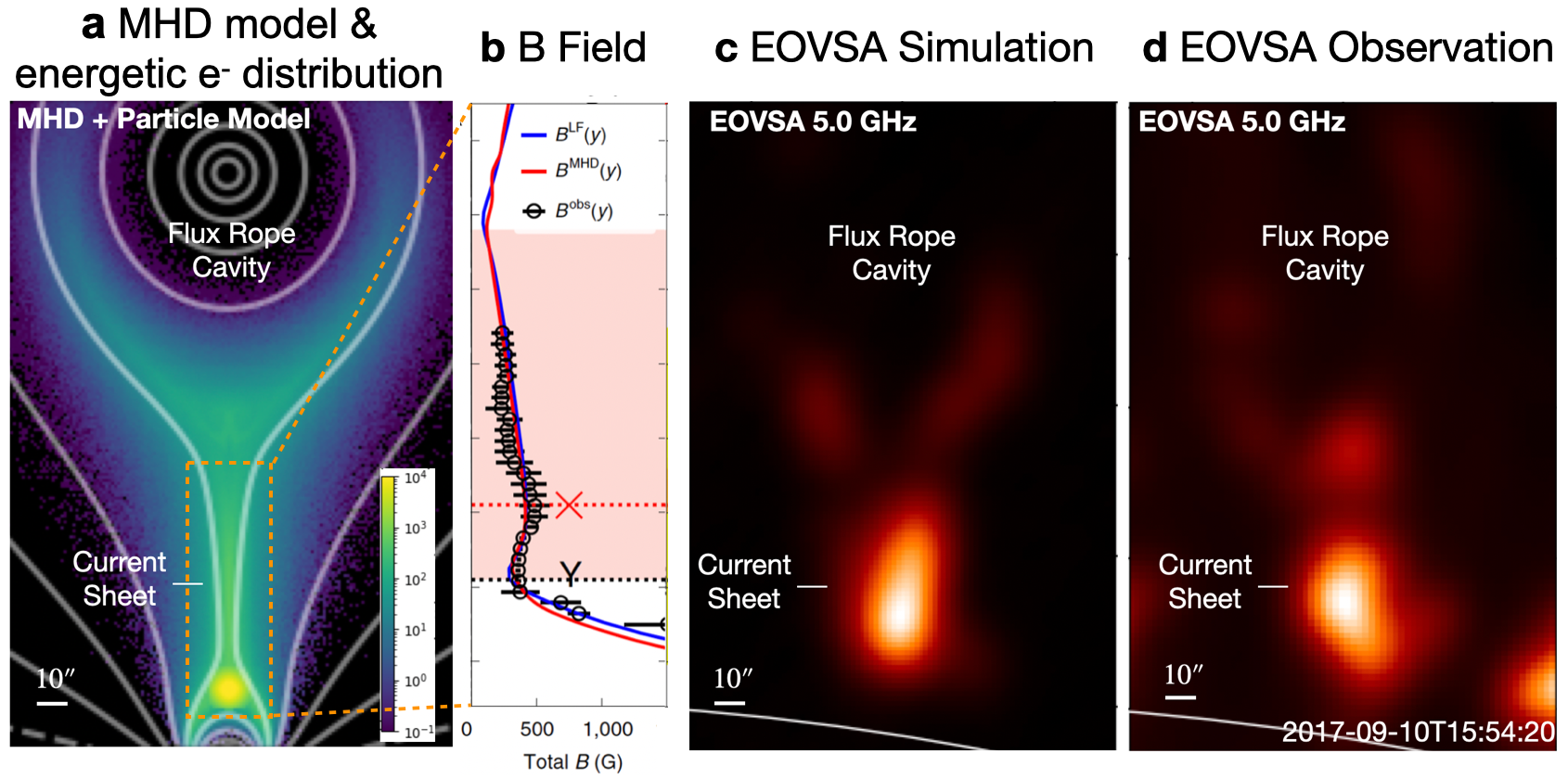}}
{\caption{Multi-wavelength observations and numerical modeling of an eruptive solar flare event on 2017 Sept. 10. The simulation is based on gyrosynchrotron radiation calculated from an input data-constrained MHD and particle model shown in (a). (b) Spatial variation of the magnetic field strength along the current sheet derived from the microwave data (black), which matches the model predictions (red/blue) \citep{Chen2020}. The simulated EOVSA image using an MHD-particle model \citep{Li2022} in (c) well resembles the actually observed EOVSA image at 5 GHz shown in (d) (see \citealt{Chen2022}). }\label{fig:eovsa}}
\end{figure}

\section{The Importance of basic theory and modeling to space weather models}

Physics-based space weather models condense our understanding of basic heliophysical processes and attempt to predict and interpret space weather events. However, in many areas of space weather modeling, the current physics models are far from mature. These applied models need constant development and input from basic theoretical and modeling studies, as the models often include important physics uncertainties. 

For example, solar energetic particle (SEP) events are thought to be generated by two processes: shocks driven by coronal mass ejections (CMEs), and magnetic-reconnection---driven acceleration in solar flares. However, their relative contributions are still not well understood \citep{Reames2020}, especially for particle acceleration associated with solar flares. For shock acceleration, the relative contributions of quasi-parallel shocks (propagation along the magnetic field) vs. quasi-perpendicular shocks remain unclear. In addition, there have been numerous events that are not consistent with one-dimensional diffusive shock acceleration theory, indicating that effects such as multi-dimensional physics, downstream turbulence, and/or reconnection may need to be included in our theoretical approach \citep{Guo2021}. 
To accurately calculate particle acceleration at near-Sun CME shocks, the sharp transition in physical properties at the shock must be resolved down to the diffusion length of energetic particles. However, multidimensional MHD CME models usually cannot resolve such scales,  while kinetic models focused on the shock front cannot reach the global scales of the full eruption. The new generation of CME-SEP models need to address these uncertainties and overcome the difficulties to achieve more reliable predictive capabilities. %Multi-moment fluid or hybrid models offer a promising way to bridge this gap, in the next decade. 

Numerical models of space weather at Earth and other planets often rely on global simulations of magnetospheric dynamics driven by interplanetary shocks and CMEs, corotating interaction regions, magnetic storms, and energy transfer to the ionosphere and thermosphere by electric fields and field-aligned currents.  The first generation of space weather forecasting tools are primarily MHD-based. However, these fluid simulations cannot robustly model key processes such as magnetic reconnection at {\it kinetic scales}, but rather rely upon resistivity or viscosity (physical or numerical) to regulate sharp structures in critically important boundary layers such as thin current sheets associated with reconnection. Such traditional proxies are crude and have limited predictive capability for space weather events. Extensive event studies over the last decade \citep[e.g.,][and the references therein]{Burch2016} have demonstrated that it is critical to include kinetic physics in the narrow reconnection layers and magnetosphere-ionosphere coupling regions.

\begin{figure}
\begin{centering}
\includegraphics[width=0.8\textwidth]{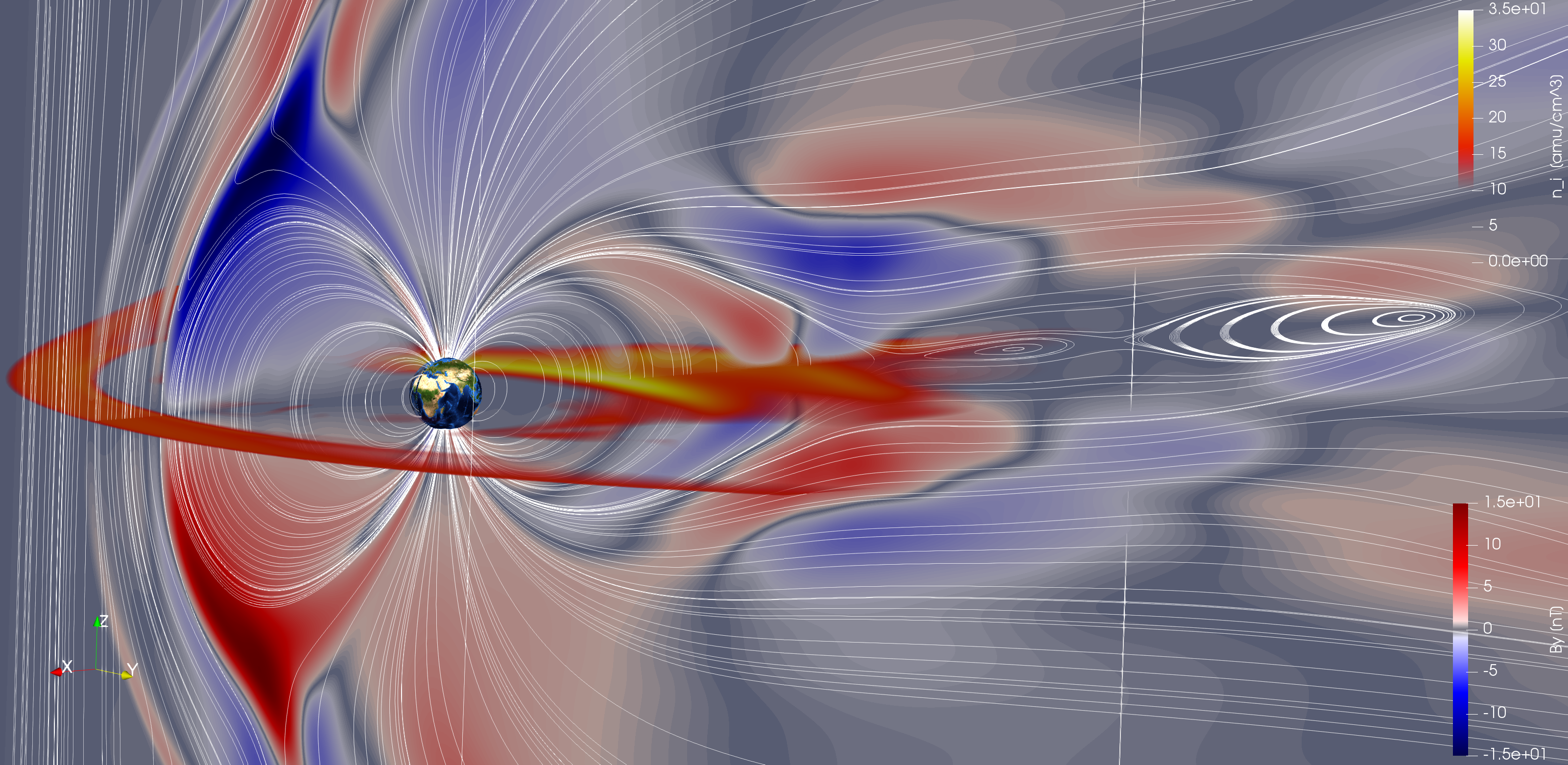}
\par\end{centering}
\caption{\label{fig:bench-earth}Perspective view of the Earth's magnetosphere from a 3D ten-moment multifluid simulation. The interplanetary magnetic field is southward. The white streamlines are magnetic field lines in the meridional plane, indicating the thinning of a current sheet on the night side, along with two flux ropes. The blue-red contours in the $xz$-plane represent the $y$ component of the magnetic field due to the Hall term in the model. The yellow-scale contours in the $xy$-plane represent ion number density \citep{Wang2020}.}
\end{figure}

These gaps have not been fully resolved and need further development. Only through persistent theoretical exploration can numerical models yield quantitative observables and predictions with enough confidence level/technology readiness level to compare with observations. There has been great progress in fundamental comprehension of particle  acceleration during magnetic reconnection \citep{Dahlin2017,Li2019,Arnold2021,Zhang2021,Li2021,Li2022}, but particle acceleration in solar flares, especially the enhanced acceleration of heavy ions and $^3$He particles, and their contributions to large SEP events, are not well understood. Moreover, the overall interaction between flares and CMEs, as well as CME-CME interactions, in producing the most energetic SEPs remain a mystery. In Earth's and planetary magnetospheres, one notable attempt to capture the physics of collisionless magnetic reconnection in the global context is the ten-moment multi-fluid model that solves the continuity, momentum, and pressure tensor equations for all ion and electron species, as well as the full Maxwell equations (\citealp{Wang2018,Dong2019,Wang2020}; see Fig.~\ref{fig:bench-earth} for its application to Earth's magnetosphere). In addition, kinetic approaches such as hybrid (fluid electrons and kinetic ions) models \citep{Karimabadi2014,Le2021} and Vlasov-hybrid models \citep{vonAlfthan2014} are promising candidates, especially by integrating kinetic electron physics, for the next generation of global magnetosphere simulations. %Conceptually it is like a fluid version of Particle-in-Cell (PIC) code, truncated at a certain order the of moment, e.g., 2nd order moment, the pressure. Non-ideal effects like the Hall effect, inertia, and tensorial pressures are self-consistently embedded without the need to explicitly solve a generalized Ohm's law. In addition, due to the fluid nature of the model, its computational cost is much cheaper than the kinetic particle codes. 

Attacking the problems outlined above with modeling advances, such as the aforementioned examples, will enhance our physical insights and provide significant theoretical inputs to space weather modeling and predictions. 

\section{The relation between theory and modeling efforts}

Theory and modeling efforts are closely related but have important differences.  Theories are developed from fundamental principles to explain poorly understood processes or phenomena, whereas modeling's goal is to apply theories to more complicated and realistic situations to compare with the ground truth of observations. They cannot replace each other and both need to be adequately supported, especially when basic theories are newly developed, and not mature enough to directly compare with observations. For example, basic studies of magnetic reconnection and associated particle acceleration often use Harris-like current sheets and simplified boundary conditions. This allows in-depth studies about basic processes and has been leading to breakthroughs  \citep[e.g.,][]{Parker1957,Loureiro2007,Guo2014,Liu2017,Cassak2017,Liu2022}.
At the same time, one often needs to take into account realistic geometries such as applications to solar eruptions,  magnetospheric storms, and coronal heating, as they require realistic 3D geometries and appropriate (e.g., line-tied) boundary conditions.  
A similar situation applies to the theory and modeling of solar eruptions. Our cartoons for the energy storage and release processes that create CMEs are still debated in part because of the extraordinary complexity of active regions observed on the sun. Similarly, the extent and kinematics of CME shocks and ejecta may be strongly influenced by the instantaneous, inhomogeneous 3D structure of the corona and inner heliosphere. Data-driven and data-constrained modeling provides an avenue to bridge this gap (often touted as the future of CME prediction and forecasting) but developing and refining such approaches require dedicated investment with a parallel focus on improving both the modeled physics (leveraging theory) and the mechanics of the approach (e.g. proper boundary conditions, data assimilation, incorporating subgrid physics, machine learning optimizations, etc).

Ultimately, it is important to recognize that theoretical understanding of many heliophysical processes is immature and incomplete, and that theory and model predictions often only address a selected set of scenarios and include significant uncertainties. Achieving closure between theory and observations, through modeling, is an iterative procedure that can take substantial effort over years. Hence continuous progress in theory and modeling is still urgently needed and federal agencies should protect against funding opportunities and review panels from negatively biasing theory/modeling proposals that do not incorporate observations. 

\vspace{-0.1cm}
\section{Recommendations}

\begin{enumerate}
\item \textbf{Restore the HTMS program to its original, inflation-adjusted level so that it meets the critical needs of heliophysics science.} More than 40 years ago, a committee led by Stirling Colgate recommended that a dedicated program be created to address large-scale, basic theory problems that required critical-mass group efforts in order to make progress, which led to a theory program now known as Heliophysics Theory, Modeling and Simulation (HTMS). The HTMS has been a huge success and has been instrumental in transforming heliophysics from a discovery-style, exploratory-science field, to a rigorous modern discipline in which closure between models and observations has become the norm. Many of the large-scale modeling groups that presently lead our field, and many of the theory/modeling leaders in our field, owe their start to the HTMS. The needs expressed in the Colgate report are even more important today, because our observations and measurements have grown enormously in depth and precision during the past decades, and require much more sophisticated theory/modeling for advancing physical understanding. The funding for the HTMS has remained constant throughout its history, however,  so that now a typical grant has only 1/3 the original buying power and is inadequate to support a critical-mass effort. We recognize that other programs have been created to tackle major unanswered questions in heliophysics, such as the NSF ANSWERS/SWQU program, the NASA LWS TR\&T, and the NASA DRIVE Centers, but none is focused on attacking basic theory and modeling problems. Furthermore, such programs support too few theory/modeling groups to make progress on the broad range of basic theory challenges that are so important for advancing heliophysics. Funding for the HTMS urgently needs to be increased by at least a factor of three, so that critical-mass groups can once again be established to attack our basic theory/modeling challenges. The original HTMS program had a transformative impact on our field, elevating it to the next level as a science discipline; a revitalized HTMS will have an equally transformative impact.

\item \textbf{Begin a program of supporting model development for next-generation basic research codes.} Heliophysics has matured to the point that large-scale numerical modeling is essential to every aspect of our science, from attacking basic theory challenges to interpreting increasingly complex observations/data and to improving space weather forecasting/nowcasting. The crucial role of modeling to meet space weather objectives has long been recognized, and the LWS Strategic Capability (SC) Program was created to respond directly to this need. The SC Program has successfully  provided many tools that the community requires for advancing space weather science. No equivalent funding program supports the development of basic research numerical tools, however, even though the need is equally pressing. Given the lack of a dedicated program, research codes are presently being developed haphazardly by cobbling together funding from multiple sources, and are not generally available to the community. Heliophysics urgently needs to develop next-generation numerical codes to attack the problems of cross-scale and cross-domain coupling that are at the heart of heliophysics science. Consequently, we recommend the creation of a Strategic Research Model program, patterned after the LWS SC Program but dedicated to funding the development of the next generation of basic-science numerical codes. All models developed by this program will have the strict requirement of being open-source and easily available to the whole heliophysics community. As part of this Strategic Research Model program, the developer groups will be supported to make the codes available through an interface such as the CCMC or GitHub, and to maintain the codes if they are sufficiently used by the community.

\item \textbf{Develop programs focused on addressing {\it mission-critical} theory and modeling needs.} As emphasized above, heliophysics is no longer a discovery-driven field; developing and advancing new understanding now drives our science. Our missions and instruments have achieved such a level of precision and sophistication that theory and modeling are integral to their success. Examples abound -- in order to connect PSP solar-wind measurements to their coronal origins, state-of-art models must be employed; in order to connect the MMS particle distribution measurements to reconnection dynamics, theory/models are essential; and in order to connect the exquisite DKIST photospheric magnetic-field observations to solar activity, again theory/models are necessary. This vital need of theory/modeling to derive science advances from the observations will only increase with future missions. No current program is focused on addressing this urgent need; consequently, we recommend the development of a dedicated theory/modeling program to accompany all future missions. These programs should begin roughly five years before mission launch and be recompeted periodically, so that the theory and models will be in hand when the data start arriving. Such mission-dedicated theory/modeling programs would increase enormously the science return from future missions, and yet have negligible cost compared to total mission funding.

\item \textbf{Develop joint theory/modeling programs with two or more basic research Agencies: NSF, NASA, and DOE.} Although instrument development and data analysis are often highly specific to the particular Agency interests, such as DOE laboratory experiments, NSF ground-based telescopes, or NASA space missions, basic theory and modeling by its very nature cut across all heliophysics science. For example, theories of magnetic reconnection and particle acceleration are essential to the science goals of all Agencies. Consequently, we recommend that joint-Agency programs be developed to address common theory/modeling needs. Some of these joint programs could include the revitalized HTMS and the  Strategic Research Model Program recommended above. Such joint programs would have the additional major benefit of maximizing the effectiveness of Agency resources. For example, the DOE has uniquely powerful resources in high-performance computing that could be made available to the NASA community as well. Furthermore, the NSF has unique strengths in educational programs, which would be of great value for building workforce infrastructure for both NASA and the DOE (see \#5). Inter-agency programs in theory/modeling would be a win-win situation for all Agencies and for heliophysics.

\item \textbf{Enhance programs for developing the next generation of theorists and modelers.} The vitality of a field is completely dependent on the strength of its community infrastructure and especially on its upcoming generation of researchers. New people bring in new ideas, and this rejuvenation is critical for making major scientific progress. The dearth of university faculty who are producing new researchers, particularly in basic theory and computation, is an issue of grave concern in heliophysics. Programs are urgently needed that address this issue, including joint programs as recommended above. A successful example is the NSF faculty development in space sciences (FDSS) program. The FDSS and similar programs are needed, but with increased frequency of solicitations and increased emphasis on basic theory/modeling. Furthermore, programs should be developed/enhanced for supporting Ph.D. students who perform their thesis research at NASA and DOE Centers in close collaboration with universities. It is essential that we develop a truly unified, inclusive heliophysics community whose workforce infrastructure integrates across science disciplines and institutions. 

\end{enumerate}

\vspace{-0.3cm}
\section{Summary}

Theory and modeling need to be independent components of heliophysics science and cannot be separated from integrated studies of heliophysics, or replaced by data analysis and data science. Without theory and modeling, sophisticated heliophysics understanding cannot be established, and empirical studies will dominate. We recommend enhanced support that encourages the development and growth of theoretical groups and young scientist communities, including but not limited to the areas discussed above. All agencies should consider expanding their current theory and modeling programs and creating new heliophysics theory programs, to cover all four main areas of heliophysics. A flagship theory and modeling program, such as a fully restored HTMS program, is critically needed. Additional new programs need to be created to develop next-generation basic research codes and address mission-critical theory and modeling needs through collaborative and coordinated efforts of basic research agencies such as NSF, NASA and DOE. We encourage the Decadal Survey Committee to identify gaps in theory and modeling support, and suggest a strategy to ensure the healthy development of theory and modeling efforts in heliophysics research. %Spacecraft missions and ground-based infrastructures should have extensive theory/modeling inputs, and space weather programs should include strong theory/modeling components. 

\newpage
%\bibliography{reconnection}

\begin{thebibliography}{}

\bibitem[Arnold et al.(2021)]{Arnold2021} Arnold, H., Drake, J.~F., Swisdak, M., et al.\ 2021, \prl, 126, 135101. doi:10.1103/PhysRevLett.126.135101


\bibitem[Bale et al.(2019)]{Bale2019} Bale, S.~D., Badman, S.~T., Bonnell, J.~W., et al.\ 2019, \nat, 576, 237. doi:10.1038/s41586-019-1818-7

\bibitem[Blandford \& Eichler(1987)]{Blandford1987} Blandford, R. \& Eichler, D.\ 1987, \physrep, 154, 1. doi:10.1016/0370-1573(87)90134-7

\bibitem[Burch et al. (2016)]{Burch2016} Burch, J. L., Moore, T. E., Torbert, R. B. \& Giles, B. L. 2016, Space Science Reviews volume 199, 5–21. doi:10.1007/s11214-015-0164-9

\bibitem[Cassak et al.(2017)]{Cassak2017} Cassak, P.~A., Liu, Y.-H., \& Shay, M.~A.\ 2017, Journal of Plasma Physics, 83, 715830501. doi:10.1017/S0022377817000666


\bibitem[Chen et al.(2020)]{Chen2020} Chen, B., Shen, C., Gary, D.~E., et al.\ 2020, Nature Astronomy, 4, 1140. doi:10.1038/s41550-020-1147-7


\bibitem[Chen et al.(2022)]{Chen2022} Chen, B. et al. \ 2022, ``Quantifying Solar Flare Energy Release'', White Paper submitted to the 2024 Solar and Space Physics Decadal Survey.

\bibitem[Cohen et al.(2022)]{Cohen2022} Cohen, I.~J. et al. \ 2022, ``The case for studying other planetary magnetospheres and atmospheres in Heliophysics'', White Paper submitted to the 2024 Solar and Space Physics Decadal Survey.

\bibitem[Crary et al.(2022)]{Crary2022} Crary, F. et al. \ 2022, ``The Magnetosphere of Jupiter: Moving from Discoveries Towards Understanding'', White Paper submitted to the 2024 Solar and Space Physics Decadal Survey.

\bibitem[Dahlin et al.(2017)]{Dahlin2017} Dahlin, J.~T., Drake, J.~F., \& Swisdak, M.\ 2017, Physics of Plasmas, 24, 092110. doi:10.1063/1.4986211


\bibitem[Dong et al.(2019)]{Dong2019} Dong, C. et al. \ 2019, Geophys. Res. Lett., 46 (21), 11584-11596. doi:10.1029/2019GL083180

\bibitem[Drake et al.(2021)]{Drake2021} Drake, J.~F., Agapitov, O., Swisdak, M., et al.\ 2021, \aap, 650, A2. doi:10.1051/0004-6361/202039432

\bibitem[Fleishman et al.(2020)]{Fleishman2020} Fleishman, G.~D., Gary, D.~E., Chen, B., et al.\ 2020, Science, 367, 278. doi:10.1126/science.aax6874

\bibitem[Garcia-Sage et al.(2022)]{Garcia-Sage2022} Garcia-Sage, K. et al. \ 2022, Star-Exoplanet Interactions: An Emerging Interdisciplinary Field in Heliophysics, White Paper submitted to the 2024 Solar and Space Physics Decadal Survey.

\bibitem[Goodrich et al.(2022)]{Goodrich2022} Goodrich, C. et al. \ 2022, ``The Persistent Mystery of Collisionless Shocks", White Paper submitted to the 2024 Solar and Space Physics Decadal Survey.

\bibitem[Green et al.(2022)]{Green2022} Green, J. L., Dong, C., Hesse, M., Young, C. A., Airapetian, V. 2022, Space Weather Observations, Modeling, and Alerts in Support of Human Exploration of Mars, White Paper submitted to the 2024 Solar and Space Physics Decadal Survey.

\bibitem[Guo et al.(2014)]{Guo2014} Guo, F., Li, H., Daughton, W., et al.\ 2014, \prl, 113, 155005. doi:10.1103/PhysRevLett.113.155005


\bibitem[Guo et al.(2021)]{Guo2021} Guo, F., Giacalone, J., \& Zhao, L.\ 2021, Frontiers in Astronomy and Space Sciences, 8, 27. doi:10.3389/fspas.2021.644354

\bibitem[Haggerty et al.(2022)]{Haggerty2022} Haggerty, C. et al. \ 2022, ``The Physics of Collisionless Dissipation in the Heliosphere", White Paper submitted to the 2024 Solar and Space Physics Decadal Survey.

\bibitem[Ji et al.(2022)]{Ji2022} Ji, H. et al. \ 2022, ``Major Scientific Challenges and Opportunities in Understanding Magnetic
Reconnection and Related Explosive Phenomena in Heliophysics and Beyond", White Paper submitted to the 2024 Solar and Space Physics Decadal Survey.

\bibitem[Karimabadi et al.(2014)]{Karimabadi2014} Karimabadi, H., Roytershteyn, V., Vu, H.~X., et al.\ 2014, Physics of Plasmas, 21, 062308. doi:10.1063/1.4882875


\bibitem[Kasper et al.(2019)]{Kasper2019} Kasper, J.~C., Bale, S.~D., Belcher, J.~W., et al.\ 2019, \nat, 576, 228. doi:10.1038/s41586-019-1813-z


\bibitem[Klimchuk et al.(2022)]{Klimchuk2022} Klimchuk, J. \ 2022, Coronal Heating, White Paper submitted to the 2024 Solar and Space Physics Decadal Survey.


\bibitem[Kollmann et al.(2022)]{Kollmann2022} Kollmann, et al. 2022, Jupiter’s radiation belts as a target for NASA’s Heliophysics Division, White Paper submitted to the 2024 Solar and Space Physics Decadal Survey.

\bibitem[Le et al.(2021)]{Le2021} Le, A., Winske, D., Stanier, A., et al.\ 2021, Journal of Geophysical Research (Space Physics), 126, e29125. doi:10.1029/2021JA029125

\bibitem[Li et al.(2019)]{Li2019} Li, X., Guo, F., Li, H., et al.\ 2019, \apj, 884, 118. doi:10.3847/1538-4357/ab4268


\bibitem[Li et al.(2021)]{Li2021} Li, X., Guo, F., \& Liu, Y.-H.\ 2021, Physics of Plasmas, 28, 052905. doi:10.1063/5.0047644

\bibitem[Li et al.(2022)]{Li2022} Li, X., Guo, F., Chen, B., et al.\ 2022, \apj, 932, 92. doi:10.3847/1538-4357/ac6efe

\bibitem[Liu et al.(2017)]{Liu2017} Liu, Y.-H., Hesse, M., Guo, F., et al.\ 2017, \prl, 118, 085101. doi:10.1103/PhysRevLett.118.085101


\bibitem[Liu et al.(2022)]{Liu2022} Liu, Y.-H., Cassak, P., Li, X., et al.\ 2022, Communications Physics, 5, 97. doi:10.1038/s42005-022-00854-x

\bibitem[Loureiro et al.(2007)]{Loureiro2007} Loureiro, N.~F., Schekochihin, A.~A., \& Cowley, S.~C.\ 2007, Physics of Plasmas, 14, 100703. doi:10.1063/1.2783986




\bibitem[Mondal et al.(2022)]{Mondal2022} Mondal, S. et al. \ 2022, ``Weak transients and the heating of the quiescent solar corona'', White Paper submitted to the 2024 Solar and Space Physics Decadal Survey.

\bibitem[Oka et al.(2022)]{Oka2022} Oka, M. et al. \ 2022, ``Particle Acceleration in Solar Flares With Imaging-Spectroscopy in soft X-rays'', White Paper submitted to the 2024 Solar and Space Physics Decadal Survey


\bibitem[Parker(1957)]{Parker1957} Parker, E.~N.\ 1957, \jgr, 62, 509. doi:10.1029/JZ062i004p00509

\bibitem[Parker(1958)]{Parker1958} Parker, E.~N.\ 1958, \apj, 128, 664. doi:10.1086/146579

\bibitem[Parker(1999)]{Parker1999} Parker, E.~N.\ 1999, \apj, 525C, 792.

\bibitem[Reames(2020)]{Reames2020} Reames, D.~V.\ 2020, \ssr, 216, 20. doi:10.1007/s11214-020-0643-5

\bibitem[Ruffolo et al.(2020)]{Ruffolo2020} Ruffolo, D., Matthaeus, W.~H., Chhiber, R., et al.\ 2020, \apj, 902, 94. doi:10.3847/1538-4357/abb594


\bibitem[Schwadron \& McComas(2021)]{Schwadron2021} Schwadron, N.~A. \& McComas, D.~J.\ 2021, \apj, 909, 95. doi:10.3847/1538-4357/abd4e6

\bibitem[Shih et al.(2022)]{Shih2022} Shih, A. et al. \ 2022, High-energy Solar Physics, White Paper submitted to the 2024 Solar and Space Physics Decadal Survey.

\bibitem[Squire et al.(2020)]{Squire2020} Squire, J., Chandran, B.~D.~G., \& Meyrand, R.\ 2020, \apjl, 891, L2. doi:10.3847/2041-8213/ab74e1


\bibitem[Sweet(1958)]{Sweet1958} Sweet, P.~A.\ 1958, Electromagnetic Phenomena in Cosmical Physics, 6, 123

\bibitem[Thorne(2010)]{Thorne2010} Thorne, R.~M.\ 2010, \grl, 37, L22107. doi:10.1029/2010GL044990


\bibitem[Turner et al.(2022)]{Turner2022} Turner, D. L. et al. 2022, Atmospheric and ionospheric impacts of energetic particle precipitation (EPP) from Earth’s ring
current and radiation belts, White Paper submitted to the 2024 Solar and Space Physics Decadal Survey.

\bibitem[von Alfthan et al.(2014)]{vonAlfthan2014} von Alfthan, S., Pokhotelov, D., Kempf, Y., et al.\ 2014, Journal of Atmospheric and Solar-Terrestrial Physics, 120, 24. doi:10.1016/j.jastp.2014.08.012


\bibitem[Wang et al.(2018)]{Wang2018} Wang, L. et al. \ 2018, J. Geophys. Res. Space Phys., 123 (4), 2815-2830. doi:10.1002/2017JA024761

\bibitem[Wang et al.(2020)]{Wang2020} Wang, L. et al. \ 2020, J. Comput. Phys., 415, 109510. doi:10.1016/j.jcp.2020.109510

%\bibitem[Wu et al.(2019)]{Wu2019} Wu, L. et al. \ 2019, Large teams develop and small teams disrupt science and technology. Nature 566, 378–382 (2019). https://doi.org/10.1038/s41586-019-0941-9

\bibitem[Zank et al.(2020)]{Zank2020} Zank, G.~P., Nakanotani, M., Zhao, L.-L., et al.\ 2020, \apj, 903, 1. doi:10.3847/1538-4357/abb828

\bibitem[Zhang et al.(2021)]{Zhang2021} Zhang, Q., Guo, F., Daughton, W., et al.\ 2021, \prl, 127, 185101. doi:10.1103/PhysRevLett.127.185101



\end{thebibliography}

\bibliographystyle{aasjournal}

\end{document}